\documentclass[prl,twocolumn]{revtex4}
\usepackage{graphicx}

\begin{document}
\title{Robust $d_{x^2-y^2}$ pairing symmetry in high-temperature superconductors}
\author{C.C. Tsuei$^1$, J.R. Kirtley$^1$, G. Hammerl$^2$, J. Mannhart$^2$, H. Raffy$^3$, and
Z.Z. Li$^3$\\}

\affiliation{$^{1}$IBM Watson Research Center, Yorktown Heights, NY, USA}
\affiliation{$^{2}$Experimental Physics VI, Center for Electronic Correlations and Magnetism,
Institute of Physics, University of Augsburg, D-86135 Augsburg, Germany }
\affiliation{$^{3}$Laboratoire de Physique des Solides, B\^{a}tement 510, UMR 8502,
Universite Paris-Sud, 91405 Orsay, France}

\date{\today}

\begin{abstract}
Although initially quite controversial, it has been widely accepted that the Cooper pairs
in optimally doped cuprate superconductors have  predominantly $d_{x^2-y^2}$ wavefunction symmetry.
The controversy has now shifted to whether the high-T$_c$ pairing symmetry changes away from optimal doping.
Here we present phase-sensitive tricrystal experiments on three cuprate systems:
Y$_{0.7}$Ca$_{0.3}$Ba$_2$Cu$_3$O$_{7-\delta}$
(Ca-doped Y-123),           La$_{2-x}$Sr$_x$CuO$_4$ (La-214) and Bi$_2$Sr$_2$CaCu$_2$O$_{8+\delta}$ (Bi-2212),
with doping
levels covering the underdoped, optimal and overdoped regions. Our work implies that
time-reversal invariant, predominantly d$_{x^2-y^2}$ pairing symmetry is robust over a large
variation in doping, and underscores the important role of on-site Coulomb repulsion
in the making of high-temperature superconductivity.
\end{abstract}

\maketitle



In the cuprate perovskites, doping with electrons or holes converts antiferromagnetic
insulators into high-temperature superconductors with a normal-to-superconducting
transition temperature T$_c$ that depends sensitively on the amount of doping \cite{orenstein}.
The doping dependence of T$_c$ can be described for many hole-doped cuprate systems
by the empirical formula \cite{presland}:
\begin{equation}
T_c(p) = T_{c,max}[1-82.6(p-p_c)^2],
\label{eq:tcvsp}
\end{equation}
where $p$ is the number of holes per CuO$_2$ layer and p$_c \approx$ 0.16.
Optimal doping represents a watershed in many superconducting
and normal state properties of the cuprate superconductors \cite{orenstein}.
Nanometer-scale charge inhomogeneity, pseudogap, and other anomalous
normal-state properties are observed mostly in the underdoped region,
while a Fermi-liquid description appears valid in the overdoped region \cite{orenstein}.
Moreover, the Hall number \cite{balakirev}, superfluid pair density \cite{panagopoulos} and many other
properties also exhibit remarkable dependencies on doping. In the language
of quantum criticality \cite{sachdev}, there are a number of competing states near
the quantum critical point ($\approx  p_c$). Variation in doping  may induce a symmetry
breaking in favor of, e.g. a spin density or charge density wave phase, or to
a superconducting phase with different pairing symmetry such as the time
reversal symmetry broken pair states $d_{x^2-y^2}+is$ or d$_{x^2-y^2} +id_{xy}$ \cite{vojta}.

Although
$d_{x^2-y^2}$ pairing symmetry is well established for several optimally doped
cuprate superconductors \cite{trirmp}, there are a number of indirect symmetry studies
suggesting a doping-induced change in pairing symmetry in some cuprates.
For example, tunnelling spectroscopy suggests a significant gapped ($s$-wave)
component in the pairing wavefunction in overdoped Y$_{1-x}$Ca$_x$Ba$_2$Cu$_3$O$_{7-\delta}$  \cite{yeh},
and a change in symmetry from $d_{x^2-y^2}$ to $d_{x^2-y^2}+ id_{xy}$  or $d_{x^2-y^2}+ is$ in
overdoped YBa$_2$Cu$_3$O$_7$ \cite{dagan}. One penetration depth measurement in electron-doped
Pr$_{2-x}$Ce$_x$CuO$_4$ indicated a $d$ to $s$ transition in the optimal to overdoped range \cite{skinta},
although a second study indicated $d$ pairing at all doping levels in this system \cite{snezko}.
In addition, the low-temperature limit of the thermal conductivity expected for a
$d$-wave superconductor \cite{lee} was observed in La$_{2-x}$Sr$_x$CuO$_4$ for all doping levels \cite{takeya}.

Phase sensitive tests of pairing symmetry using the tricrystal geometry have
been described elsewhere \cite{trirmp}. Briefly, thin films of cuprate superconductors
are epitaxially grown on a SrTiO$_3$ substrate composed of 3 grains. The substrate
and subsequent cuprate thin film geometry are chosen (Fig. 1(a)) such that,
for a $d_{x^2-y^2}$ superconductor, in the absence of supercurrents there are an odd
number of sign changes in the component of the pairing wavefunction normal
to the grain boundaries upon circling the tricrystal point. The sign changes
at the grain boundaries
cost Josephson coupling energy. This energy is reduced by the
generation of circulating supercurrents, resulting in a Josephson vortex
with a vortex number N$_\phi$ =1/2. For real superconducting order parameters and
the conventional I$_c$=I$_1$sin($\phi$)  Josephson current-phase relationship the tricrystal
vortex should have exactly half of the conventional flux quantum
$\phi_0$=h/2e=2.07$\times$10$^{-15}$ Wb of magnetic flux. In our geometry (Fig. 1(a)),
a spontaneous N$_\phi$ =1/2 vortex will occur at the tricrystal point only if
the $d_{x^2-y^2}$ component is at least as large as a possible $s$ component.
Further, a fixed imaginary component to the order parameter would make
the absolute value of the flux in the tricrystal point vortex different
upon inversion by roughly the ratio of the imaginary component to the
total pairing amplitude. The N$_\phi$ =$\pm$1/2 vortices at the tricrystal point
could have the same absolute values of flux, if there were domains with
imaginary order parameter components, but such domains would have to
alternate signs on a length scale smaller than the experimental spatial
resolution.  In our experiments, the flux at the tricrystal point is
imaged with a scanning Superconducting Quantum Interference Device (SQUID)
microscope (SSM). In our SSM, the sample is scanned using a mechanical
lever mechanism relative to a well shielded pickup loop integrated into
a SQUID sensor \cite{ssmapl}.
\begin{figure}
\includegraphics[width=3.5in]{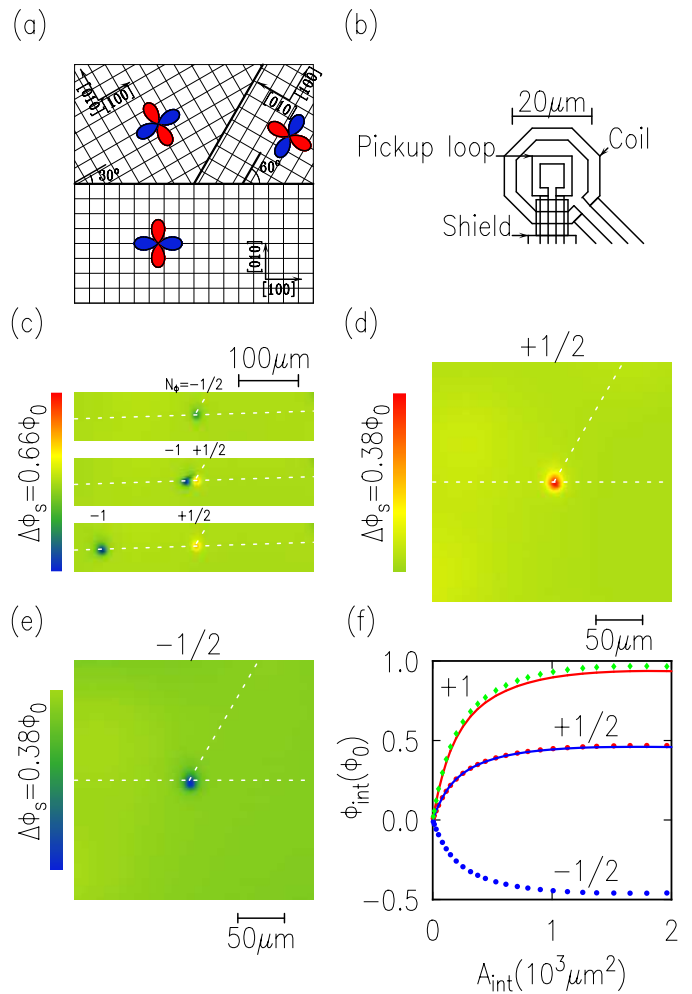}
\vspace{0.1in}
\caption{(a) Tricrystal geometry. The polar plots represent the pairing
wavefunctions, with the red lobe phases shifted by $\pi$ relative to the blue.
(b) Schematic of the pickup loop area of the SQUID susceptometer sensor used.
(c,d,e) are SQUID microscope images of the tricrystal point of a
Y$_{0.7}$Ca$_{0.3} $Ba$_2$Cu$_3$O$_{7-\delta}$ film
on a SrTiO3  tricrystal with the geometry of (a). All SSM images in this paper
were taken at 4.2K. The dashed lines indicate grain boundaries. (c)
illustrates the inversion of a N$_\phi$ =-1/2 vortex at the tricrystal point
(top image). (c, middle image) a 5mA pulse of current is passed through
the susceptometer field coil to invert a -1/2 vortex (c, top image) to a
+1/2 vortex, creating also a N$_\phi$  = -1 Josephson vortex in the horizontal
grain boundary. (c, bottom image) the -1 Josephson vortex is dragged
from the tricrystal point by moving the sensor parallel to the grain
boundary while applying a current of 4mA. The color scales span $\phi_s$ = 0.66
$\phi_0$ of flux through the SQUID pickup loop in (c) and 0.38 $\phi_0$ in (d) and (e).
(f) Shows the integration of the total flux (in units of  $\phi_0$) of the N$_\phi$  = +1/2 state
(d, red dots), the -1/2 state (e, blue dots), and a nearby N$_\phi$  = +1 integer vortex
(green dots) over a circular area A$_{int}$ centered at the tricrystal point.
The blue line in (f) is the N$_\phi$ =-1/2 data multiplied by -1, demonstrating
time reversal invariance. The red line in (f) is the N$_\phi$  = +1/2 data multiplied by 2.
}
\label{fig:dopesc1p}
\end{figure}

Fig. 1(c,d,e) shows SSM images of N$_\phi$ =$\pm$1/2 vortices at the tricrystal point of a
130nm thick
Ca-doped Y-123  film
epitaxially grown by pulsed laser deposition on a SrTiO$_3$ tricrystal in the
geometry of Fig. 1a, chosen to show the half-flux quantum effect for a
superconductor with $d_{x^2-y^2}$ pairing symmetry.
These films have sufficiently high grain boundary
supercurrent densities \cite{hammerl} that Josephson vortices in the grain boundaries
and at the tricrystal point are resolution limited with a 4 micron diameter
pickup loop. The N$_\phi$ =+1/2 state can be inverted to its degenerate time-reversed
-1/2 state (Fig. 1(c)) by applying currents of a few mA with the appropriate
polarity through the field coil of the SQUID susceptometer \cite{gardner}.

Integration of the total flux at the tricrystal point (Fig. 1(f)) shows that
the N$_\phi$  = $\pm$1/2 states have the same total flux and field distribution. The time
reversal symmetry broken pair states such as $d_{x^2-y^2} +id_{xy}$ and $d_{x^2-y^2} +is$ are
thus ruled out.

In contrast to Ca-doped Y-123, La-214 films grown on SrTiO$_3$ tricrystals have
relatively low T$_c$'s and very small grain boundary critical current densities,
presumably due to the lattice mismatch between the deposited film and the substrate \cite{si}:
The fields of the half-fluxon at the tricrystal point for the T$_c$=28.5K sample shown in
Fig. 2 spread out along the grain boundaries over tens of microns. Nevertheless,
it is clear that there are spontaneous currents at the tricrystal point, and that
the resulting vortex retains time- reversal symmetry (Fig. 2(b)).
For detailed modelling, the flux/unit length in the ith branch of the vortex can be written as (18)
\begin{equation}
\frac{d\phi_i}{dr_i} = \frac{\phi_0}{2\pi}\frac{-4a_i}{\Lambda_{Ji}}\frac{e^{-r_i/\Lambda_{Ji}}}{1+a^2_ie^{-2r_i/\Lambda_{ji}}}
\label{eq:flux}
\end{equation}
where $r_i$ is the absolute value of the distance along the $i$th grain boundary from the
tricrystal point,
$J_i$ is the Josephson penetration depth of the ith grain boundary, $a_i/(\Lambda_{Ji}(1+a^2_i))$
is the same for each grain boundary, and the total flux at the tricrystal point is
$\phi=\sum_i(\phi_0/2\pi)4\tan^{-1}(a_i)$.
The flux through the SQUID pickup loop is calculated by propagating the fields at
the surface to the height of the pickup loop using Fourier transform techniques,
and then integrating over the known pickup loop geometry. Results of fits of these
calculations to the data (Fig. 2(c)) show that to within our experimental errors,
the vortex at the tricrystal point has a total flux of $\phi_0/2$.

\begin{figure}
\includegraphics[width=3.5in]{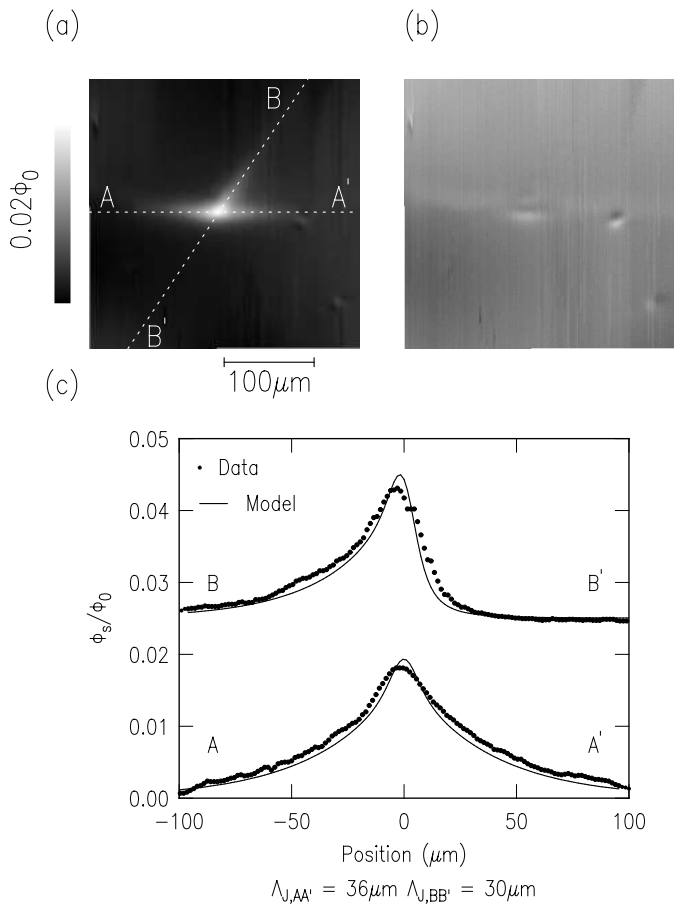}
\vspace{0.1in}
\caption{
SQUID microscope images, taken with an 8$\mu$m square pickup loop,
of the tricrystal point region for an underdoped (T$_c$=28.5K, thickness=310nm)
La-214 film epitaxially grown by laser ablation.  (a) Difference image
($\phi_s$(+20nT)-$\phi_s$(-20nT))/2 (to cancel out stray fields)
for images taken with the sample cooled in $\pm$20nT
fields, resulting in N$\phi$ = $\pm$1/2  Josephson
vortices at the tricrystal point. The sum image ($\phi_s$(+20nT)+$\phi_s$(-20nT))/2
(b) shows that the flux from the N$_\phi$  = $\pm$1/2 vortices cancel out within a
few percent, demonstrating time-reversal symmetry.  (c) Cross-sections
through the data of (a) along the lines indicated by the arrows, and
fits using the Josephson penetration depths $\Lambda_J$ along the two grain
boundaries as fitting parameters, assuming the vortex at the tricrystal
point has $\phi_0/2$ of magnetic flux. The best fit value, allowing the total
flux at the tricrystal point to vary, was  $\phi$=0.585$\pm$0.1 $\phi_0$.}
\label{fig:lscofig}
\end{figure}

\begin{figure}
\includegraphics[width=3.4in]{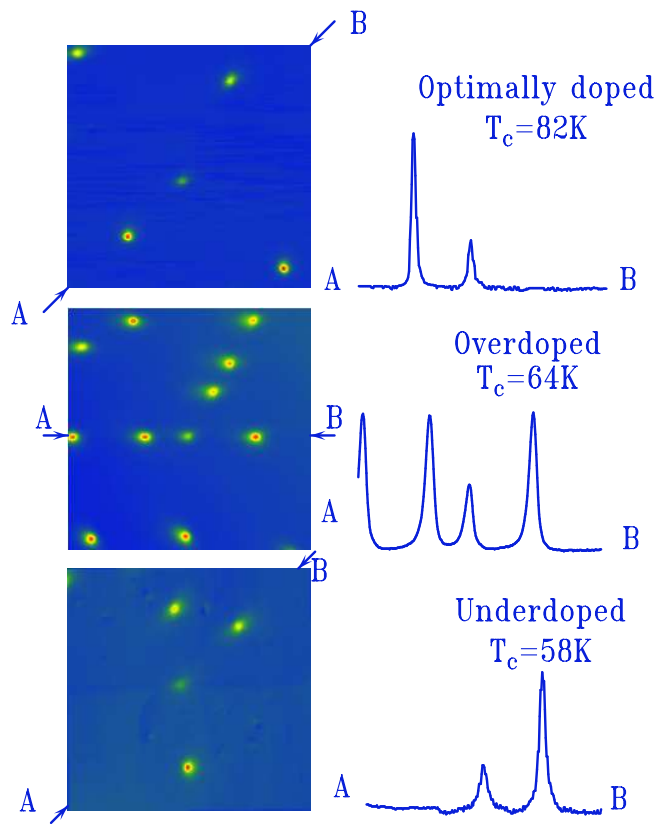}
\vspace{0.1in}
\caption{SQUID microscope images for tricrystal samples with optimally,
underdoped, and overdoped Bi$_2$Sr$_2$CaCu$_2$O$_{8+\delta}$  300nm thick films epitaxially
grown by RF sputtering on a SrTiO$_3$ substrate with the geometry of Fig. 1(a).
The images were taken with a 4$\mu$m octagonal pickup loop (optimal) and a 7.5$\mu$m
square pickup loop (overdoped and underdoped). Full scale variation in  $\phi_s$
was 0.4$\phi_0$, 0.18$\phi_0$, and 0.21 $\phi_0$ respectively. The lines are cross-sections
of the image data through the tricrystal point along the directions indicated.
In each case there is a N$_\phi$  = +1/2 vortex at the tricrystal point.
}
\label{fig:bsccdope}
\end{figure}

Tricrystal experiments were also performed on a number of Bi-2212 superconductors
with various doping levels achieved by controlling the oxygen content during film
deposition, or thermal annealing of a given tricrystal Bi-2212 film \cite{konstantinovic} (Fig. 3).
Modelling like that described above \cite{blnktprl} showed that in all cases the amount of magnetic flux at the
tricrystal point was $\phi_0$/2 to within experimental error, and that the absolute
value of the total flux at the tricrystal point remained the same within a
few percent when the vortex was inverted, either by applying a local field
at low temperature, or by cooling again in a slightly different field. The
crystal structure of all the Bi-2212 samples studied here is tetragonal
equivalent \cite{trirmp}. A $d+s$ mixed pair state in Bi-2212 is therefore symmetry forbidden.

Fig. 4 displays the doping range covered in this study. Numerous studies
indicate that the crystal structure in multi-grain cuprates is preserved
except in a narrow region 1nm in width along the grain boundaries \cite{gbjrmp}.
It is possible that the dopant concentration at the grain boundary is
different from the bulk. However, our temperature dependent scanning
susceptibility measurements on Y$_{0.7}$Ca$_{0.3}$Ba$_2$Cu$_3$O$_{7-\delta}$ bicrystals indicate
that the supercurrent across the grain boundary has a T$_c$ within 5K of
that in the bulk, and it is unlikely that the dopant profile near the
grain boundary remains constant while the grains themselves undergo a
full range of variation. Therefore, our results show that in several
cuprate systems, over a wide range of doping, pairing with $d_{x^2-y^2}$
symmetry is robust, and that any fixed imaginary component of the order
parameter in pair states such as  $d_{x^2-y^2} +id_{xy}$ and $d_{x^2-y^2} +is$ must be
small. Note that $d_{x^2-y^2}$ pairing symmetry in the Bi-2212 system persists
to $p \approx$0.07 (Fig. 4), very close to the onset of superconductivity at $p$=0.05.
At such low doping other orders such as anti-ferromagnetism, charge density
wave and spin density wave phases compete vigorously with $d$-wave superconductivity \cite{sachdev,vojta}.

\begin{figure}
\includegraphics[width=2.0in]{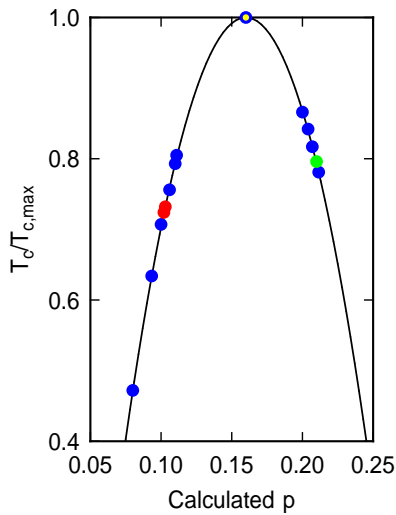}
\vspace{0.1in}
\caption{
Plots of T$_c$/T$_{c, max}$ vs. $p$, the calculated doping holes per CuO$_2$ layer, using Eq. 1,
with $p_c$ = 0.16, and T$_{c, max}$ = 90K, 90K, 38K and 82K for optimally doped
YBa$_2$Cu$_3$O$_{7-\delta}$  (yellow dot), Y$_{0.7}$Ca$_{0.3}$Ba$_2$Cu$_3$O$_{7-\delta}$
(green dot), La$_{2-x}$Sr$_x$CuO$_4$ (red dots)
and Bi$_2$Sr$_2$CaCu$_2$O$_{8+\delta}$  (blue dots) respectively.
}
\label{fig:dopetcvp}
\end{figure}

As Anderson \cite{anderson87,anderson04} first suggested, the physics of both the normal and
superconducting states of the cuprates is dictated by strong Coulomb interactions
in the CuO$_2$ planes, although the exact form of the ground-state wavefunction is
still a matter of debate \cite{varma}. Many theoretical studies including strong correlation
effects favour $d_{x^2-y^2}$ over $d_{xy}$ and extended $s$-wave ($s^*$) states \cite{dagotto,bulut}, with pairing
in the $d_{x^2-y^2}$ channel enhanced by on-Cu-site repulsion and suppressed by inter-site
Coulomb interactions \cite{plekhanov}. The stability of the pure $d_{x^2-y^2}$ pair state is further
enhanced by the presence of the van Hove singularity or flat band around (0,$\pi$), ($\pi$,0)
in the 2D band structure of the CuO$_2$ planes \cite{trirmp,dagotto}. The origin of such Fermi surface
pinning can be attributed to the effect of on-site repulsion \cite{himeda}.

The robust nature of $d_{x^2-y^2}$ pairing over a wide doping range (0 $< p \leq$   0.35) has been
demonstrated by several numerical studies based on Hubbard models \cite{paramekanti,sorella}, consistent with our present work.
Furthermore, our observation of d$_{x^2-y^2}$ pairing in the low doping regime $p \approx$ 0.07 is
supported by a recent measurement of c-axis penetration depth as a function of temperature
and doping in YBa$_2$Cu$_3$O$_{7-\delta}$, suggesting that the $d_{x^2-y^2}$ nodal quasi-particles survive to
very low doping \cite{hosseini}.

The present work, coupled with previous work establishing $d$-wave symmetry for a number
of optimally hole- and electron-doped cuprates \cite{trirmp}, calls for a universal origin of
$d_{x^2-y^2}$ pairing symmetry in all the cuprate superconductors studied so far. We suggest
that all evidence points to a strong influence of strong on-site Coulomb repulsion, a
characteristic common to all cuprates, and which is also responsible for the doping-induced
Mott insulator-metal transition observed in all cuprate perovskites.

The work of G.H. and J.M. was supported by the BMBF (13N6918), the DFG (SFB484)
and the ESF (PiShift). We would like to thank R.H. Koch and D.M. Newns for useful discussions.



\end{document}